\documentclass[preprint,proceedings]{rmaa}
\usepackage{paralist}

\usepackage{psfrag,color}

\newcommand\chb{$C({\rm H}\beta)$}
\newcommand{\grsim}{\mathrel{\hbox{\rlap{\hbox{\lower4pt\hbox{$\sim$}}}\hbox{$>$}}}}

\SetYear{2003}
\SetConfTitle{Energetics of Cosmic Plasmas}

\title{Photoionization Models of Metal-poor Extragalactic \\
\ion{H}{ii} Regions and the Primordial Helium Abundance}

\author{
  V. Luridiana,\altaffilmark{1} 
  A. Peimbert,\altaffilmark{2}
  and M. Peimbert\altaffilmark{2}
}

\altaffiltext{1}{Instituto de Astrof\'\i sica de Andaluc\'\i a (CSIC), Granada, Spain.}
\altaffiltext{2}{Instituto de Astronom\'\i a (UNAM), M\'exico D.F., Mexico.}

\shortauthor{Luridiana et al.}
\shorttitle{Metal-poor extragalactic \ion{H}{ii} regions}

\fulladdresses{
\item V. Luridiana: Instituto de Astrof\'\i sica de Andaluc\'\i a (CSIC), 
c/ Camino Bajo de Hu\'etor 24, Apartado 3004, 18080 Granada, Spain
  (\email{vale@iaa.es}).
\item A. Peimbert and M. Peimbert: Instituto de Astronom\'\i a (UNAM), 
Ap.do postal 70-264, 04510 M\'exico D.F., Mexico
(\email{antonio, peimbert@astroscu.unam.mx}).
}

\listofauthors{V. Luridiana, A. Peimbert, M. Peimbert, \& M. Cervi\~no}

\indexauthor{Luridiana, V.}
\indexauthor{Peimbert, A.}
\indexauthor{Peimbert, M.}

\abstract{We discuss the effects of collisional enhancement of Balmer lines
on the determination of the primordial helium abundance.
To this aim, we present a photoionization model of the metal-poor 
extragalactic \ion{H}{ii} region {SBS~0335--052}.
We show that the derived helium abundance ($Y$) of this \ion{H}{ii} region
depends on the amount of collisional excitation affecting the Balmer lines,
both directly (through an underestimation of the actual He/H ratio)
and indirectly (through an overestimation of the interstellar reddening).
We detail how each of these effects affects the derived value of $Y$.}

\resumen{Discutimos el efecto de la contribuci\'on colisional a la
intensidad de 
las l\'\i neas
de Balmer en la determinaci\'on de la abundancia de helio primordial.
Con este objetivo, presentamos un modelo de fotoionizaci\'on
de la regi\'on \ion{H}{ii} extragal\'actica de baja metalicidad 
{SBS~0335--052}.
Mostramos que la abundancia de helio ($Y$) derivada en esta regi\'on
depende de la contribuci\'on colisional a las l\'\i neas
de Balmer, tanto directa (a trav\'es de una subestimaci\'on del
cociente He/H real) como indirectamente (a trav\'es de una 
sobreeestimaci\'on de la extinci\'on interestelar).
Desglosamos la forma en que ambos efectos afectan a la determinaci\'on 
de $Y$.
}

\addkeyword{galaxies: abundances}
\addkeyword{galaxies: ISM}
\addkeyword{H~II regions}
\addkeyword{ISM: abundances}

\begin{document}
\maketitle

\section{Introduction}
\label{sec:intro}

This contribution highlights the difficulties
of obtaining accurate determinations of the
helium abundance ($Y$) in low-metallicity \ion{H}{ii} regions.
High-quality $Y$ values are necessary
in order to reduce the uncertainty in the
derived value of the primordial helium abundance ($Y_P$),
which is a fundamental quantity in cosmology:
see also the reviews by \citet{L03} and
\citet{P03m} for more complete discussions of this topic 
and a quantitative estimate of the error budget in the determination of $Y_P$.

One of the main contributors to the overall uncertainty
in $Y$ for individual regions
is the collisional enhancement of Balmer lines.
In \ion{H}{ii} regions, the main process
contributing to the intensity of Balmer lines is
H$^+$ recombination,
but a further contribution can arise from
the collisional excitation of H$^0$
from the ground state.
This contribution is generally a minor one,
but it might reach a value of several percent
under appropriate conditions; since it depends both
on the fraction of H$^0$ and on the temperature,
its evaluation requires both these quantities to be
known accurately.

To illustrate the difficulties inherent to this task
and suggest possible solutions,
we present a photoionization model of
{SBS~0335--052}, an extremely
low-metallicity extragalactic \ion{H}{ii} region,
and show how the abundance analysis of this object
should be modified to take into account the collisional enhancement.
In a forthcoming paper \citep{Lal03} we will present
a more extended analysis of this object,
as well as photoionization models of two other
low-metallicity regions, {H 29} and {I Zw 18}.
See also \citet{DK85}, \citet{SK93}, \citet{SI01}, and \citet*{PPL02}
for previous discussions of this topic.

Low-metallicity regions are the most affected
by collisions, since the collisional excitation of Balmer lines
depends very steeply on temperature,
which is especially high in these objects.
As a rule of thumb, collisional contribution may be 
safely neglected in the analysis of objects with $T_e \lesssim 15,000$~K,
while in hotter objects it 
may enhance the H$\beta$ intensity 
by up to several percent.
If not taken into account, this enhancement has two biasing effects 
on the determination of $Y_P$:
first, if the observed intensity is interpreted in terms 
of pure H$^+$ recombination,
the He$^+$/H$^+$ abundance is underestimated;
second, 
since the percentual collisional increase in H$\alpha$ 
is always higher than that in H$\beta$,
the measured H$\alpha$/H$\beta$ ratio is larger 
than it would be without collisions,
producing a spurious reddening in the Balmer spectrum
\citep{FO85}.
If this extra-reddening is mistaken for the effect of
interstellar extinction,
the dereddened ratios $I(\lambda)/I({\rm H}\beta)$
derived for lines blueward of H$\beta$ are overestimated, 
and those derived for lines redward of H$\beta$ are underestimated.
The net effect of such bias is a decrease of the derived $Y$
value, 
since the helium lines redward of H$\beta$ are globally
brighter than those blueward of H$\beta$
and weigh more in the abundance analysis.
We will show this effect quantitatively in 
the case of {SBS~0335--052}, the second most metal-poor 
\ion{H}{ii} region known.

\section{{SBS~0335--052}\label{sec:sbs0335}}

{SBS~0335--052} is a roundish region with 
a linear radius $R\sim 800$ pc.
The reddening coefficient \chb\ determined by \citet{Ial99}
varies from 0.225 to 0.33 along the slit; 
we will adopt in the following the value $C({\rm H}\beta) = 0.25$,
corresponding to the total $F({\rm H}\alpha)/F({\rm H}\beta)$ in the slit.

The model presented in this paper was computed 
with the photoionization code Cloudy 94 
\citep{F00a,F00b}; the ionizing source was computed 
with Starburst99 \citep{Lal99}.
We adopted the following input parameters:
$Q({\rm H}^0) = 2.2 \times 10^{53}$ s$^{-1}$;
$Z_{*}=0.0010$;
$t=3.2$ Myr;
$Z_{gas}=0.0007$;
filling factor $\epsilon = 1.9\times 10^{-3}$;
covering factor $cf = 0.80$;
and a Gaussian density law
$N_{\rm e} = {\rm max}\,(N_{\rm e}^{\rm max} {\rm exp}(r/r_0)^{-2}, N_{\rm e}^{\rm min})$, 
with $N_{\rm e}^{\rm max} = 600$ cm$^{-3}$, $N_{\rm e}^{\rm min} = 90$ cm$^{-3}$,
and $r_0 = 225$ pc.
The results will be compared against the observational
data by \citet{Ial99}.
The model's output is presented both for the complete
Str\"omgren sphere, and modified to take into account
the aperture effect introduced by the $1''\times 5.4''$ slit
used by  \citet{Ial99}, which was divided into
 9 extractions (Figure~\ref{fig:slits}).
The model's predictions will be presented
in the following figures
as a function of the radius of the structure,
both in linear (parsec) and angular (arcsec) units;
the bottom axis is in arcsec for those plots
that simulate observational (projected) quantities,
and in parsecs for those plots
that describe theoretical (radial) quantities.

\begin{figure}[!t]
 \includegraphics[width=\columnwidth]{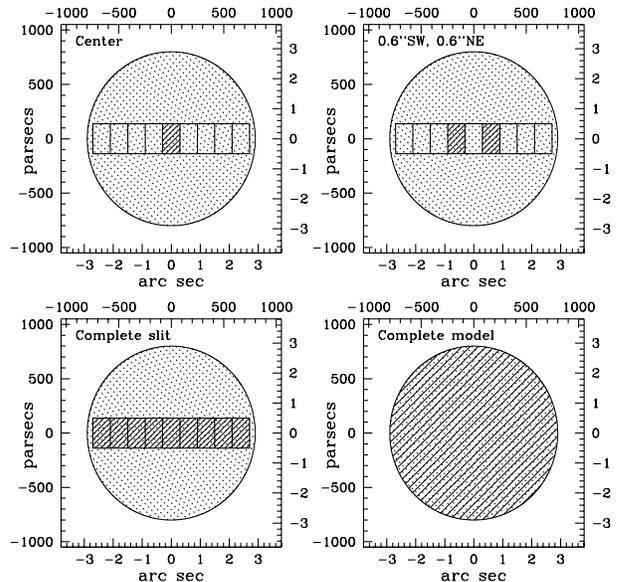}
 \caption{Sketch of the slit used by \citet{Ial99},
 and the nine extractions making it up, projected onto
the model nebula (see text). 
 The assumed distance is $d=57$ Mpc.
 \label{fig:slits}}
\end{figure}

\begin{figure}[!t]
 \includegraphics[width=\columnwidth]{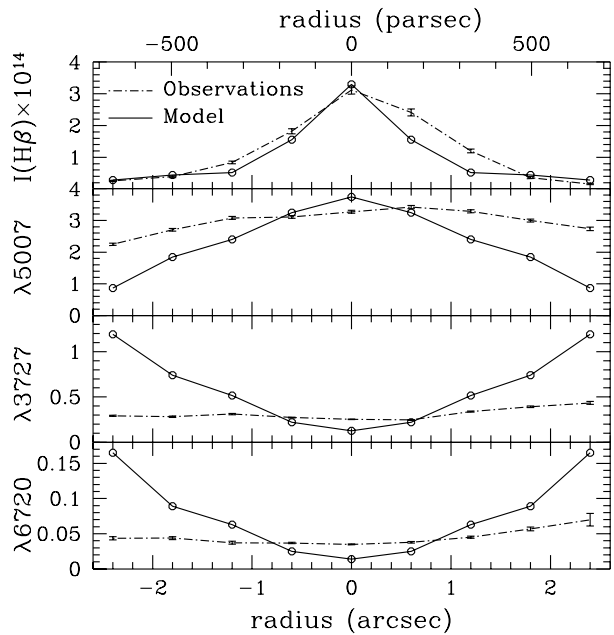}
 \caption{Predicted profile of some emission lines (solid line)
compared to the observational data (dot-dashed line).
 \label{fig:profile_lines}}
\end{figure}

Figure~\ref{fig:profile_lines} shows the profiles 
of the principal emission lines predicted by our model.
The theoretical points correspond to the simulation of 
the nine extractions making up the slit; 
they are compared to the intensities observed by \citet{Ial99}.

\begin{figure}[!t]
 \includegraphics[width=\columnwidth]{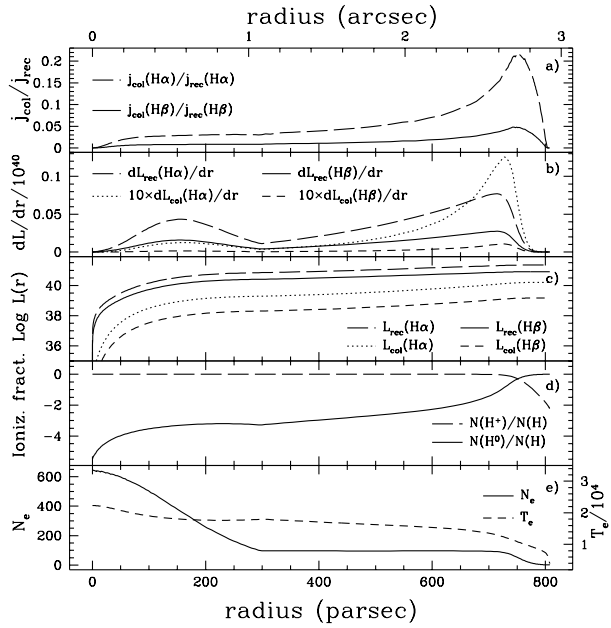}
 \caption{Predicted radial behavior of several quantities
related to collisional enhancement:
a) ratio between 
the collisional and the recombination emissivities of
H$\alpha$ and H$\beta$; b) differential contribution
of the collisional and the recombination process to the
H$\alpha$ and H$\beta$ luminosities;
c) cumulative collisional and recombination 
H$\alpha$ and H$\beta$ luminosities;
d) hydrogen ionization structure;
e) electronic temperature and density.
 \label{fig:profile_L}}
\end{figure}

\begin{figure}[!t]
 \includegraphics[width=\columnwidth]{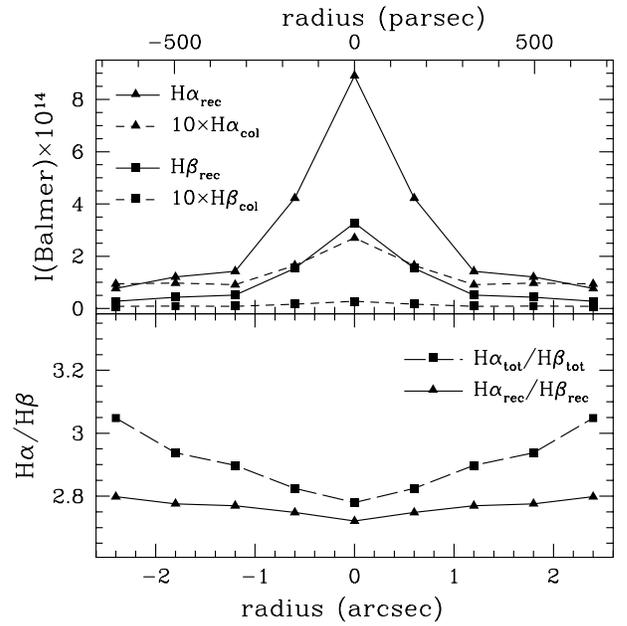}
 \caption{Top: projected profiles of the H$\alpha$ and H$\beta$ luminosities,
with the aperture effect taken into account;
bottom: comparison between the predicted
profile of the total H$\alpha$/H$\beta$ ratio
and the profile of the H$\alpha$/H$\beta$ ratio that would results
from pure recombination. 
 \label{fig:HaHb}}
\end{figure}

Figure~\ref{fig:profile_L} shows the radial (theoretical)
behavior of various quantities related to the collisional
contribution: from the top down, the ratio between 
the collisional and the recombination emissivities of
H$\alpha$ and H$\beta$; the differential contribution
of the collisional and the recombination process to the
H$\alpha$ and H$\beta$ luminosities;
the cumulative collisional and recombination 
H$\alpha$ and H$\beta$ luminosities;
the hydrogen ionization structure;
and, in the bottom panel, the electronic temperature and density.
The top panel of Figure~\ref{fig:HaHb} compares the 
projected profiles of the H$\alpha$ and H$\beta$ luminosities,
with the aperture effect taken into account;
the bottom panel of the same figure compares the predicted
profile of the total H$\alpha$/H$\beta$ ratio
to the profile of the H$\alpha$/H$\beta$ ratio that would result
from pure recombination. 
The recombination profile is almost constant,
as expected from recombination theory for a structure
with a fairly constant temperature profile;
however, the collisional contribution, which peaks
around $R\sim 700$ pc as a result of the interplay 
between the temperature and the H$^0$ fraction
(Figure~\ref{fig:profile_lines}),
increases the total H$\alpha$/H$\beta$ ratio along the
whole diameter, and particularly in the outermost extraction
(which is centered on $R\sim 600$ pc).

\begin{figure}[!t]
 \includegraphics[width=\columnwidth]{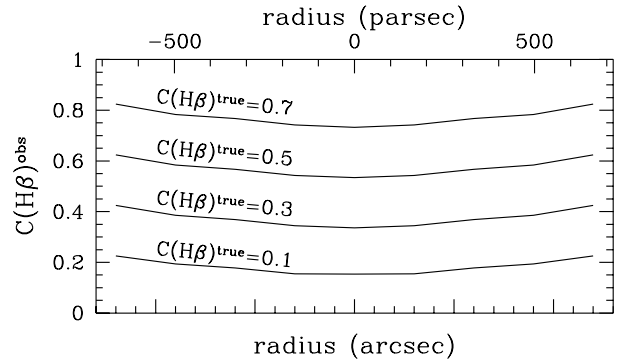}
 \caption{Predicted reddening coefficient  along the model nebula
for various assumption on the true extinction coefficient
$C({\rm H}\beta)_{\rm real}$.}
 \label{fig:cHbeta}
\end{figure}

This increase in the total H$\alpha$/H$\beta$ ratio
would be observationally undistinguishable
from an increase due to interstellar extinction:
in Figure~\ref{fig:cHbeta} we give examples of
the reddening coefficient which would be observed
along the nebula
for various assumptions on the true extinction coefficient
$C({\rm H}\beta)^{\rm true}$.

\section{Effects of collisions on the determination of $Y$}\label{sec:collisions}

Table~\ref{tab:SBShelium} presents the values of some helium line fluxes, 
along with three sets of reddening-corrected line intensities.  
In the first set we neglect the collisional effects 
in the hydrogen lines; 
in the second we subtract the collisional contribution 
from the observed $I({\rm H}\beta)$, but we neglect 
the collisional effects on the observed reddening;
finally, in the third we subtract the collisional contribution from both
$I({\rm H}\beta)$ and $C({\rm H}\beta)^{\rm obs}$.
The $C({\rm H}\beta)$ values associated
to each intensity set are also listed.  
The table also presents the resulting $y^+$ values,
and the difference in the derived helium abundance $\Delta Y$
for each of the three cases;
these numbers are based on a very simple analysis,
and are meant to provide only a rough indication of collisional effects.
A more sophisticated analysis, based on better models and 
on an improved algorithm for abundance analysis, will be presented
in \citet{Lal03}.

The following facts should be considered when
reading this table:
first, the effect depends on the number of lines used;
second, the lines redward of H$\beta$ are more brilliant,
and therefore dominate the abundance determination;
third, the effect might be a strong function of the position,
especially in real nebulae, in which the behavior 
of the temperature and ionization degree are much less smooth
than in a model.

\begin{table}[!t]\centering
  \setlength{\tabnotewidth}{\columnwidth}
  \tablecols{5}
  \setlength{\tabcolsep}{1.2\tabcolsep}
  \caption{Helium line intensities and abundances\tabnotemark{a}}\label{tab:SBShelium}
  \begin{tabular}{lrccc}
   \toprule
\multicolumn{1}{c}{{$\lambda$}} &\multicolumn{1}{c}{{$\frac{F(\lambda)}{F({\rm H}\beta)}$}}  
&\multicolumn{1}{c}{{$\frac{I(\lambda)}{I_{tot}({\rm H}\beta)}$}}&\multicolumn{1}{c}{{$\frac{I(\lambda)}{I_{rec}({\rm H}\beta)}$}}
&\multicolumn{1}{c}{{$\frac{I(\lambda)}{I_{rec}({\rm H}\beta)}$}}\\
   \midrule
4026 & 1.294  & 1.473 & 1.490  & 1.461 \\
4471 & 3.404  & 3.616 & 3.659  & 3.626 \\
4921 & 0.821  & 0.816 & 0.825  & 0.826 \\
5876 & 11.476 & 10.169 & 10.290 & 10.477  \\
6678 & 3.237  & 2.631 &  2.662  & 2.746  \\
{H$\alpha$} & {3.339} &&&\\
   \midrule
{$C({\rm H}\beta)$}&& {0.25} & {0.25} & {0.21}\\
$\langle y^+ \rangle$&& {0.0825}& {0.0835} & {0.0846} \\
$\Delta Y$&& {0.0000}& {0.0023} & {0.0047} \\
    \bottomrule
    \tabnotetext{a}{Observed \ion{He}{i} line fluxes, relative to H$\beta$,
and corresponding dereddened intensities for three different
assumptions on collisional effects (see Section~\ref{sec:collisions}).}
\end{tabular}
\end{table}

\section{Summary and conclusions}

In this work we discuss the influence of collisional enhancement
of hydrogen lines on the accuracy of $Y_P$ determinations.
The final goal of this project, which is presented
in its entirety in \citet{Lal03},
is to quantify the correction that should
be applied to previous $Y_P$ determinations in which
collisional enhancement of Balmer lines has been neglected,
and to estimate the uncertainty attached to such correction.
In this framework, we also investigate to which extent the predicted
collisions are model-dependent,
and note that the occurrence 
of temperature fluctuations in real regions would enhance 
the collisions with respect to predicted values.

We show that the helium abundance in individual objects
may have been underestimated as a consequence of the H$\beta$ collisional
enhancement.
The underestimation in the $Y$ values arises from
the combination of two factors:
the collisional H$\beta$ enhancement
and the collisional reddening.
Since the last factor mimics a higher $C({\rm H}\beta)$,
the quantitative effect of the collisional bias on $Y_P$ 
is a function of the \ion{He}{i} lines
used in the analysis.

As a consequence of this bias, the $Y_P$ value has also been underestimated.
Since the effect on the individual $Y$ values is more pronounced at
high electronic temperatures, which characterize low-metallicity objects,
the net effect on the $Y - Z$ relation is a decrease of the slope.

\acknowledgments

VL is supported by a Marie Curie Fellowship
of the European Community programme {\sl ``Improving Human Research Potential 
and the Socio-economic Knowledge Base''} under contract number HPMF-CT-2000-00949.
This project has been partially supported by the AYA 3939-C03-01 program.

\end{document}